\documentstyle[11pt,newpasp,twoside,epsf]{article}
\def\edcomment#1{\iffalse\marginpar{\raggedright\sl#1\/}\else\relax\fi}
\reversemarginpar

\newcommand\mj{M$_{Jupiter} $}

\begin{document}

\title{Companions to Young Stars}
\author{Patrick J. Lowrance}
\affil{Jet Propulsion Laboratory, Infrared Processing and Analysis Center, 
Pasadena, CA 91125}
\author{\& the NICMOS Environments of Nearby Stars team \& STIS 8176 team}

\begin{abstract}
Brown dwarfs occupy the important region in the mass range 
between stars and planets. Their existence, ambigious until only recently, and their 
properties give insight into stellar and planetary formation. We present 
statistical results of 
an infrared, coronagraphic survey 
of young, nearby stars that includes probable companions to three 
young G-type stars, Gl 503.2 (G2V), HD 102982 (G3V), and Gl 577 (G5V).
The companion to Gl 577 is a possible binary brown dwarf, according to evolutionary models. 
A dynamical determination of the components' masses 
will be achievable in the near future and be an excellent 
test of the predictive ability of the evolutionary models.
\end{abstract}

\section{Introduction}


To explore fully the 
similarities and differences between stellar and planetary formation, 
the study of brown dwarfs as companions is essential. Since the 
primary's age and distance have been determined, these two properties, 
usually uncertain for field brown dwarfs, are already known. 
The primary goal of this survey of 45 young (t$<$300Myr), nearby (d$<$50pc)
stars was the discovery of substellar companions. Because substellar objects 
are hotter and brighter when younger, they can be more easily detected. 
Each target star was selected based on its being single and 
having at least two youth criteria (chromospheric activity, lithium, proper motion) 
indicating a young age. The final sample had a median age of 150 Myr 
and a median distance of 30 pc. 
The survey utilized the infrared coronagraph on the  
Near-Infrared Camera and Multi-Object Spectrometer (NICMOS) on the Hubble 
Space Telescope (HST). Earlier results include the brown dwarf companions 
TWA5B (Lowrance et al 1999; L99) and HR 7329B (Lowrance et al 2000). 

\section{Observations with NICMOS on the Hubble Space Telescope}

The observing strategy was to place each star
behind the coronagraph, observe for approximately 800s, roll the telescope by 30 degrees, 
and observe again for 800s (actual integration times were
adjusted to fit in variable orbit times). Following the method 
fully described in L99, when we subtract the two images, 
a positive and negative image of any candidate companion should remain.

\section{Determining Detection Limits}

To determine our sensitivity limits within the NICMOS 
roll-subtracted images of the observed stars, we 
planted\footnote{software courtesy of A. Ghez and A. Weinberger.} 
point-spread-function (PSF) stars, 
generated with Tiny Tim (Krist \& Hook 1997), 
at random locations in every image. 
These PSF stars are noiseless, can be
adjusted in flux, and were stepped by 0.2 mag to 
examine a range of magnitudes from H=10--22 for each subtraction.

\begin{figure}[t]
\plottwo{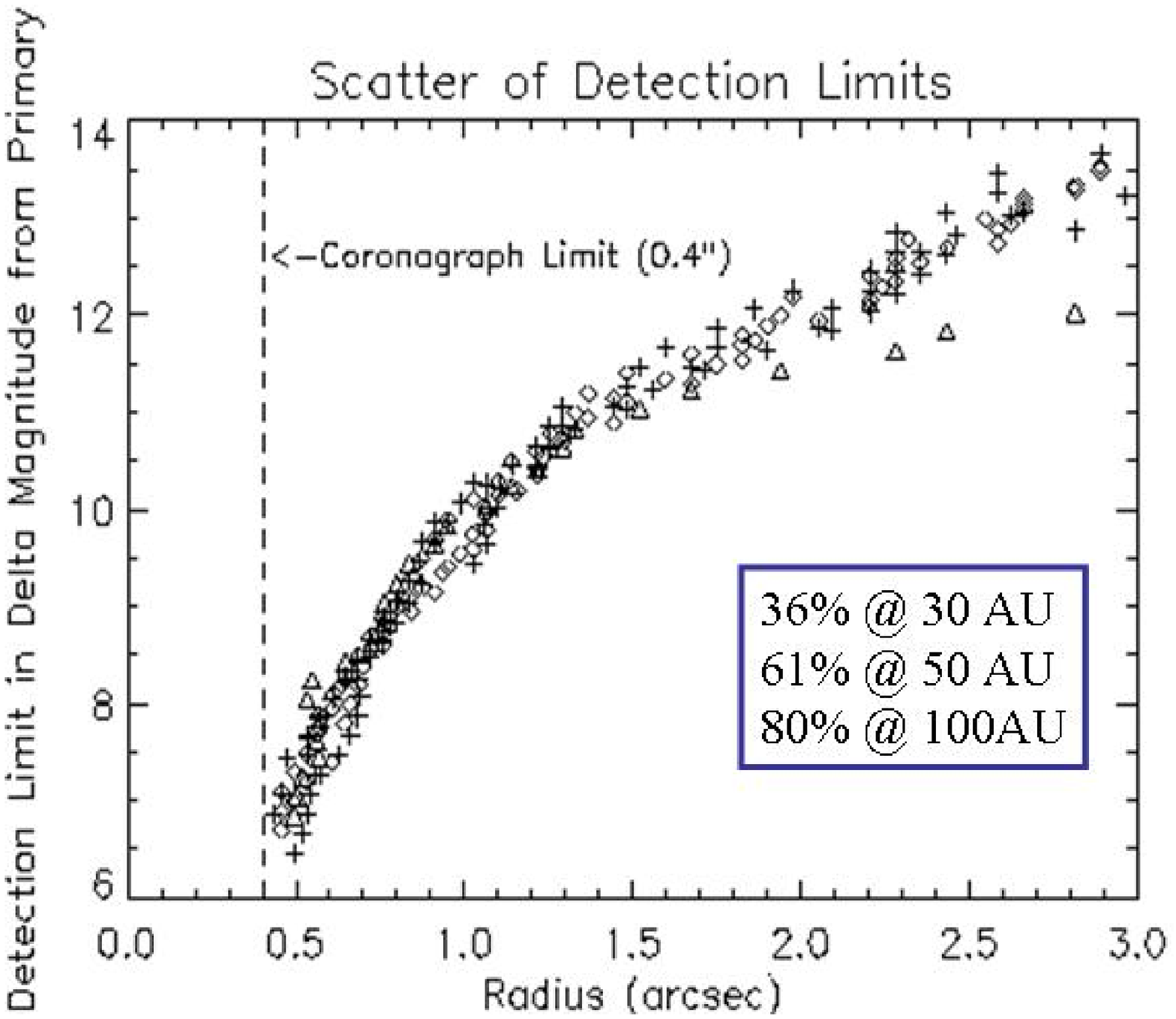}{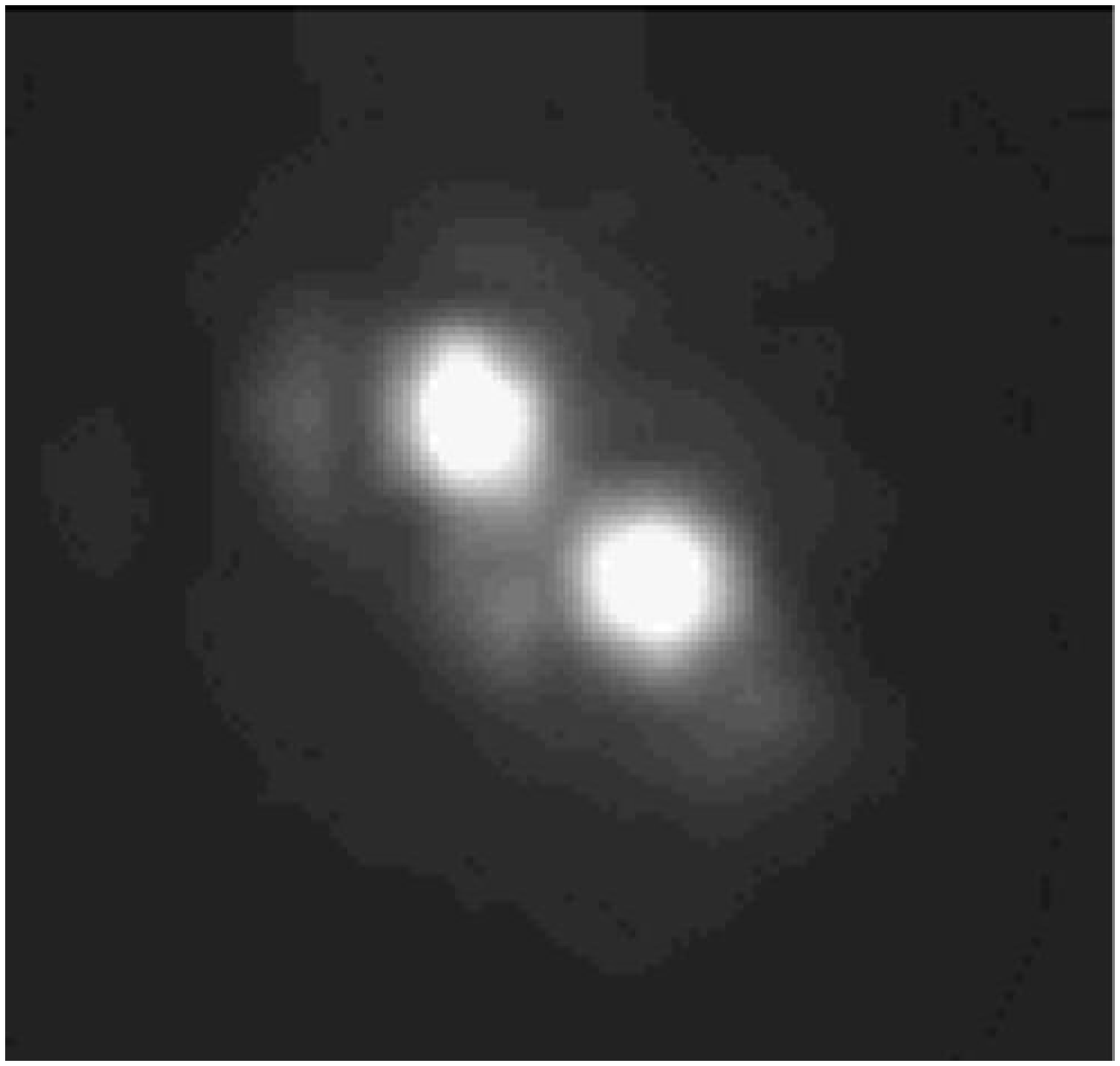}\label{scatter}
\caption{(a)Detection limits reached in survey of young stars. Inset box indicates 
percentage of  objects around which a 5 M$_{Jupiter}$ object would have been detected. 
Different symbols indicate the different brightness of the primary star plotted; diamonds 
for H$=$5 mag to triangles for H$=$9 mag.(b) The Adaptive Optics image of the companions to 
Gl 577A, clearly indicating a binary companion with separation 0.082$\arcsec\pm$0.005 that 
corresponds to $<$4 AU at that distance.} \label{scatter}
\end{figure}

\begin{figure}
\plottwo{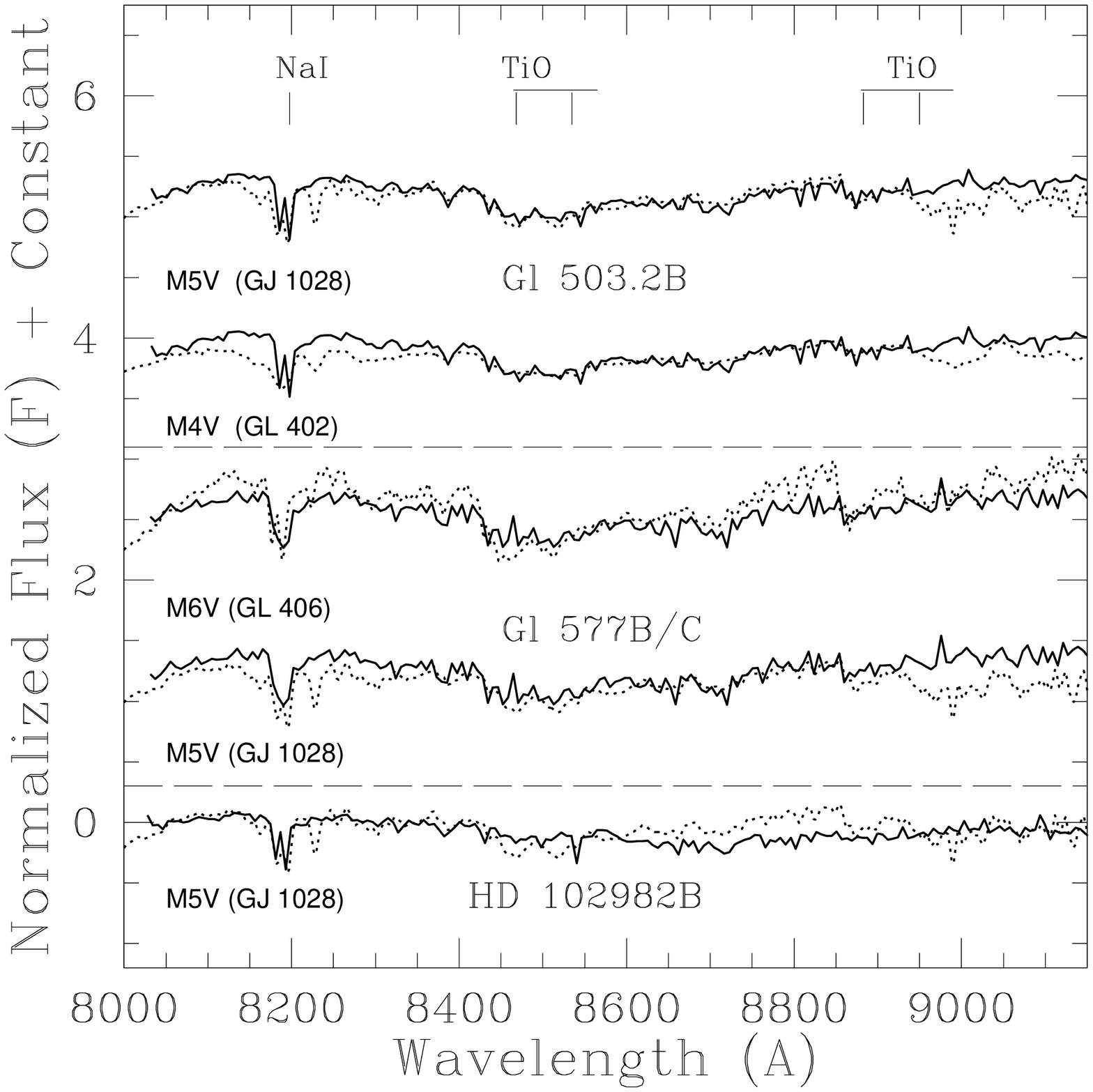}{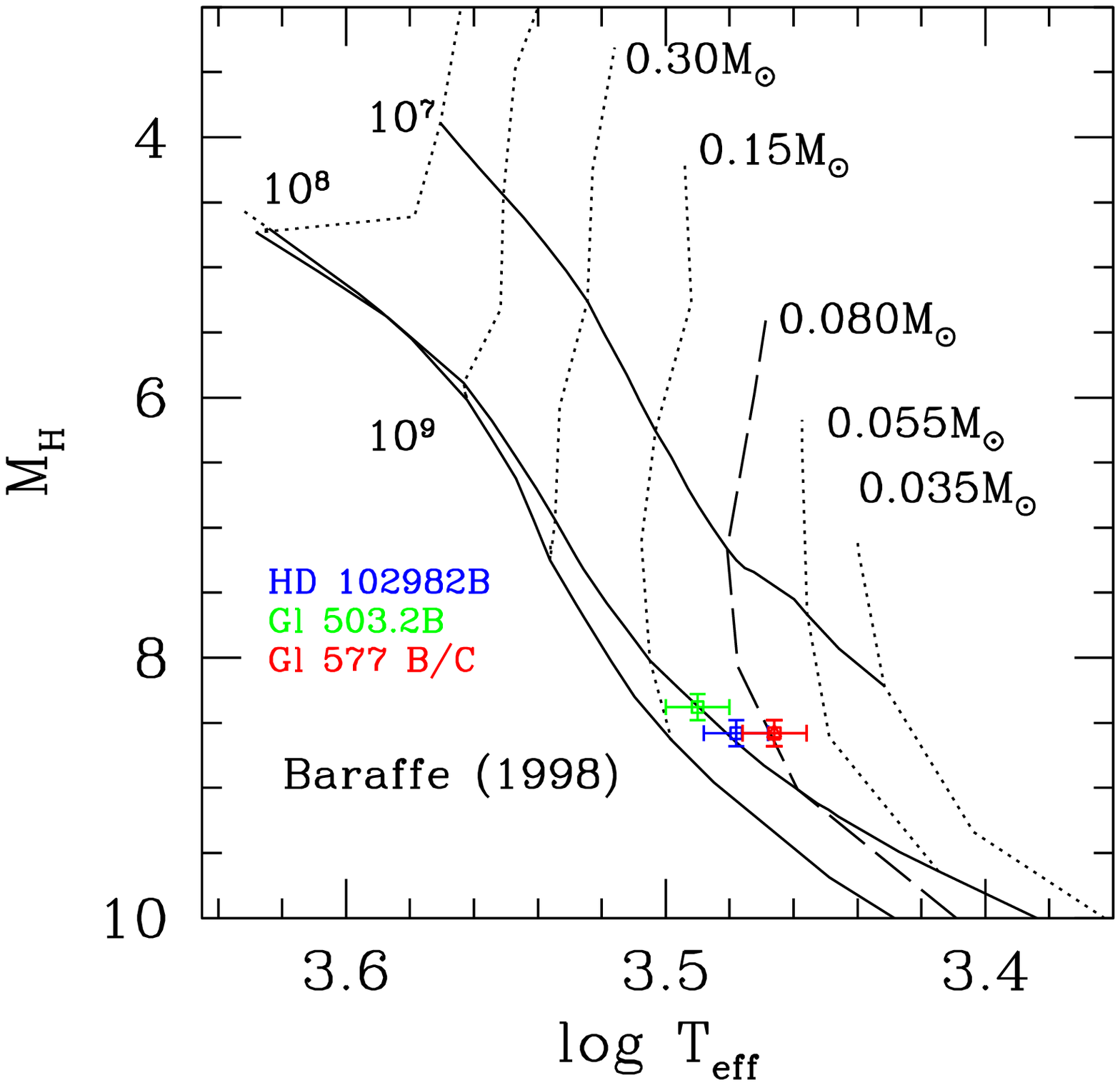}\label{double}
\caption{(a)STIS spectra of HD 102982, Gl 577B/C, and Gl 503.2B  (solid)
(normalized from ergs/s/cm$^2$/\AA) separated by the dashed lines,
compared with standard late-type M dwarf spectra (dotted)
(Kirkpatrick et al. 1991;Kirkpatrick et al. 1997). 
The zero level of each
spectrum is -1.2, 0, 1.3, 2.7 and 4 respectively)
(b)Evolutionary diagrams with all three companions to the G stars. The derived ages of the secondaries 
match the primaries' ages implied from youth criteria. The binarity of 
Gl 577 B/C has been taken into account, and it lies on the stellar/substellar boundary.} \label{double}
\end{figure}

In figure 1, we plot the detection limits found overall in
the observations. 
Our sample has an average primary magnitude H $=$ 7 mag and a median age of 0.15
Gyr. At 1$\arcsec$, we can detect $\Delta$H$=$9.5 mag for all 
stars. At a median
distance of 30 pc, 1$\arcsec$=30AU and our average limit corresponds
to M$_H$ = 14.1 mag. From
models (Burrows, A., personal communication), this corresponds to less than 20 \mj\ for the median age. 

This program fully sampled 45 young stars with the ability to detect 30 \mj\ 
brown dwarfs at 0.5 $\arcsec$. For some 
separations from the primary, we were able to detect objects into the high
mass planet range. For our oldest stars, t=0.3 Gyr, a 5 \mj\ object is
expected to have a absolute H mag of 18.7 mag (Burrows, A., personal
communication). Closer than 5$\arcsec$, detectability depends on the brightness, age, and distance 
of the primary. We find that a 5 \mj\ object could have been 
detected around 61\% of the primaries at 50AU (Figure 1a), approximately the outer edge 
of the solar system.

\section{Follow-up Observations of Candidate Companions}

We followed up the candidate companions to Gl 503.2, 
HD 102982, and Gl 577 
with spectra from the HST Space Telescope Imaging Spectrograph (STIS) and 
using Adaptive Optics (AO) imaging at several ground-based telescopes.  

All three secondaries were observed with the STIS. 
Each primary was acquired into the 
52$\arcsec\times$0.$\arcsec$2 slit and offset based on the NICMOS
astrometric results to place the secondary into the slit. 
Spectral imaging sequences were
completed in one orbit per star with the G750M grating in three tilt settings
executing a two position dither along the slit at each. 

The STIS spectra were calibrated, averaged, binned to a resolution of
$\sim$ 6\AA, and normalized to the flux (in ergs/s/cm$^2$/\AA) at 8500\AA. 
We compared the final spectra to those of standard star spectra (see Figure 2a). 
The best fit appears to lie between
M4V and M5V for Gl 503.2B, between M5V and M6V for Gl577B/C and 
M5V for HD 102982B. 

The Gl 577 and Gl 503.2 systems were observed 
with the Canada-France-Hawaii telescope (CFHT) on March 4 and 5, 
1999 (UT) using the AO system PUEO (Rigaut et al. 1998) and 
with the 200 inch Hale telescope at Palomar Observatory on May 14, 
2000 (UT) using the AO system PALAO. 
The Gl 577 system was also observed on 12 Aug 2000 using the AO 
system on the 10-m W.M. Keck II telescope (Wizinowich et al.
2000) with the slit-viewing camera (SCAM) on NIRSPEC (McLean
et al. 1998). 
Basic reduction included correction for bad pixels and flat-fielding procedures. 

A common proper
motion between the primaries and secondaries was established. 
Both the Gl 577A and Gl 503.2A proper motions 
have been measured accurately by the Hipparcos (Perryman et
al. 1997) satellite. 
The proper motion for HD 102982 is listed in the 
Tycho catalogue (Hog et al. 2000). From adaptive optics observations, 
the Gl 577 and Gl 503.2 systems 
were each measured three times. The HD 102982 system was measured in the STIS observation.   
All were within 1 sigma of the previously measured separation 
and 3 sigma outside the predicted separation from proper motion. We therefore conclude
that Gl 503.2B, HD 102982B, Gl 577B/C are companions to the brighter primaries.

\section{Derived Mass and Age}
From their placement on pre-main sequence evolutionary tracks (Baraffe
et al. 1998), we can infer a mass for the secondaries (Figure 2b). 
HD 102982B and Gl 503.2B appear to be consistent
with 100 Myr low-mass ($<$ 0.15 M$_{\sun}$) stars. 
Gl 577 B/C appear to lie at the theoretical
stellar/substellar border mass of 0.08 M$_{\sun}$ from the infrared magnitude and
spectral type. Both components are of equal magnitude, so 
both might be brown dwarfs.

There are very few stars of mass $<$ 0.2 M$_{\sun}$ with a dynamical
determination of their mass, as few short-period
binaries in this mass range are resolved. High 
spatial resolution provided by HST and adaptive optics now makes 
it possible to resolve fainter binaries with smaller separations and 
shorter periods. With a separation of only $\sim$4 AU, 
a dynamical mass for Gl 577BC can be 
determined in less than a decade. Only then will the 
predictive ability of evolutionary models be tested near 
the stellar/substellar boundary. 
Further infrared adaptive optics observations along 
with radial velocity measurements are currently underway.

\end{document}